\documentstyle[aps,12pt]{revtex}

\begin{document}
\title{The 3d-electron states in {\it \ }LaCoO$_3$}
\author{Z. Ropka}
\address{Center for Solid State Physics, S$^{nt}$ Filip 5, 31-150 Krak\'{o}w,%
}
\author{R.J. Radwanski}
\address{Center for Solid State Physics, S$^{nt}$ Filip 5, 31-150 Krak\'{o}w.%
\\
Inst. of Physics, Pedagogical University, 30-084 Krak\'{o}w, Poland.\\
email: sfradwan@cyf-kr.edu.pl.}
\maketitle

\begin{abstract}
Three fundamentally different electronic structures for 3d electron states
in LaCoO$_3$, discussed in the current literature, have been presented. We
are in favour of the localized electron atomic-like approach that yields the
discrete energy spectrum associated with the atomic-like states of the Co$%
^{3+}$ ions in contrary to the continouos energy spectrum yielded by band
theories. In our atomic-like approach the d electrons form the
highly-correlated system 3d$^n$ described by S=2 and L=2.

PACS\ No: 71.70.Ej : 75.10.Dg : 73.30.-m;

Keywords: Mott insulators, electronic structure, crystal field, LaCoO$_3$
\end{abstract}

\date{(3.08.2000)}

LaCoO$_3$ belong to the class of compounds known as Mott insulators. It
appeals the scientific interest by more than 50 years.The uniqueness of LaCoO%
$_3$ is mostly related with its non-magnetic ground state at low
temperatures and with the significant violation of the Curie-Weiss law at
low temperatures [1,2]. Most of Co compounds are magnetic, even strongly.
Despite of enormous long lasting theoretical efforts the description of LaCoO%
$_3$ is still under very strong debate. The fundamental controversy ''how to
treat the d electrons'' starts already at the beginning - should they be
treated as localised or itinerant. Directly related with this problem is the
structure of the available states: do they form the continuous energy
spectrum like it is in the band picture, schematically shown in Fig. 1, or
the discrete energy spectrum typical for the localised states.

The aim of this paper is to show up the fundamentally different descriptions
of LaCoO$_3$ appearring in the current literature by presenting 3 different
electronic structures for the 3d electrons. One is related with the band
description and two with the localised description. There is a number of
band-structure calculations [3-9] and the results are schematically shown in
Fig. 1. The bands are not polarised with respect to the spin direction as
LaCoO$_3$ does not order magnetically down to lowest temperatures. The same
occupancy of the spin up and down states reproduces the
experimentally-observed non-magnetic (diamagnetic) state at 0 K and is the
realisation of the so-called low-spin (LS)\ state.

In the current literature, apart of the band picture, often appears a
localized electronic structure like presented in Fig. 2 [10, 4]. At low
temperatures the low-spin state (LS) is realized for 6 d electrons (spins)
that is characterized by S=0. The electrons (spins) are put subsequently one
by one on the single-electron states formed for 1 d electron in the
octahedral crystal field: lower triplet t$_{2g}$ and higher doublet e$_{g}$.
With increasing temperatures the LS state transforms to the high-spin (HS)
state with S=2. Korotin et al. [4] pointed out the existence of the
intermediate spin (IS)\ state with S=1 at the middle temperature region. The
temperature-induced transformation of the LS state to the HS state has been
originally inferred in Refs 1 and 2 from the temperature dependence of the
paramagnetic susceptibility that exhibits an intriguing maximum at about 90
K.

The general shape of the bands presented in Fig. 1 can be understood knowing
the localised states of Fig. 2. The continuous energy spectrum looks like
the smooth convolution on the available localised single 3d electron
orbitals t$_{2g}$ (occupied) and higher e$_g$ orbitals (empty). The
similarity of the band density of states and the energy level scheme of Fig.
2 is related with the single-electron treatment of 3d electrons in both
approaches.

Recently more complex electronic structure has been derived within the
localized picture [11,12] that is presented in Fig. 3. The authors of Refs
11-12 have pointed out that the 6 d electrons form the highly-correlated
atomic-like 3d$^6$ system. Two Hund's rules yield the ground term $^5$D with
S=2 and L=2. This term is 25-fold degenerated. Its degeneracy is removed by
the crystal field and spin-orbit interactions. Under the action of the
dominant cubic crystal field, the $^5$D term splits into the orbital triplet 
$^5$T$_{2g}$ and the orbital doublet $^5$E$_g$ with the energy separation
about 2-3 eV. In the octahedral oxygen surrounding, realized in the
perovskite structure of LaCoO$_3$, the orbital triplet $^5$T$_{2g}$ is
lower. The low-energy electronic structure has been calculated from the
single-ion-like Hamiltonian considered within the 25-fold LS space
containing simultaneously the crystal-field and spin-orbit coupling. The
trigonal off-cubic distortion of the octahedral site, relevant to the
situation realized in LaCoO$_3$, produces the non-magnetic singlet (in the $%
\left| \text{LSL}_{\text{z}}\text{S}_{\text{z}}\right\rangle $ space) ground
state and two excited doublets that turn out to be highly magnetic, see
Fig.3. These excited states become thermally populated with the increasing
temperature - as these states are strongly magnetic their thermal population
manifests in the susceptibility experiment as a maximum that can be
misleadingly interpreted as a temperature induced low-high spin transition.

It is worth notinh that in our calculations we get the low- and high-spin
state within the one term $^{5}$D (for it, the intra-atomic spin-orbit
coupling is essentially important [13]) in contrary to the two terms (LS,
HS) or three terms (LS, HS, IS) considered within the single-electron
approach of Fig. 2. Moreover, in the energy level scheme shown in Fig. 3 one
can find the origin for the intermediate-spin state - the first and the
second excited states of Fig. 3 one can try to describe by the effective
spin of 1.16 and 1.83, respectively - these values are accidentally close to
the ad hoc assumed values of S=1 and S=2. These states are in the discussed
energy interval, up to 80 meV. The possibility of getting the non-magnetic
ground state, discussed in the literature as the low-spin state, as well as
the intermediate and highly-magnetic states within the same term we take as
the great plus for our atomic-like approach. Surely the explanation
involving one term only is physically simpler to be realized - according to
the well-known Occam's razor principle the simpler explanation is the better
one.

We are in favour of the atomic-like approach. It enables the calculations of
whole termodynamics in remarkably good agreement with experimental results
[11,12]. The diamagnetic state of LaCoO$_{3}$ is associated with the
non-magnetic singlet ground state of the Co$^{3+}$ ion realized in the
high-spin state (S=2) in the presence of the orbital magnetism and the
intra-atomic spin-orbit coupling. This non-magnetic Co$^{3+}$ state differs
significantly from the low-spin nonmagnetic state with the term $^{1}$A$_{1}$%
, the LS configuration in  Fig. 2. Such low-spin state is thought in the
current literature to be realized by the strong crystal field (strong CEF\
approach). Our nonmagnetic state is found within the weak crystal-field
regime, i.e. when the crystal field does not break the intra-atomic
arrangement about the total S and L, thus preserving the atomic structure.
This weak crystal-field approach, known within the rare-earth CEF\ community
as the CEF approach, has been successfully applied to 3d-ion doped systems,
when 3d ions are introduced as impurities [14-17]. Such systems have been
successfully studied in electron paramagnetic resonance (EPR) experiments
[14-17]. In Refs. 11 and 12 we have applied this approach to a 3d-ion system
where Co ions are the part of the solid, LaCoO$_{3}$ in this case.

We would like to point out that our approach should not be considered as the
treatment of an isolated ion - we consider the cation in the octahedral
crystal field. This octahedral crystal field is predominantly associated
with the oxygen octahedron CoO$_6$. The perovskite structure is built up
from the corner sharing octahedra CoO$_6$ - thus such the atomic structure
occurs at each cation due to the translational symmetry. The strength of the
crystal field interactions is determined by the whole charge surroundings,
not only by the nearest oxygen octahedron. It makes that the CEF approach
looks like a single-ion approach but in fact describes coherent states of
the whole crystal.

Within the single-electron approaches, both the band theory and the strong
CEF approach, the realization of the LS and HS state results from an
interplay of the CEF and exchange interactions ($\Delta _{CEF}$ and J$_{ex}$
in Fig. 2). The different strength of CEF and exchange interactions in
different compounds is reasonable but the postulated their significant
change with temperature in LaCoO$_{3}$ is, according to us, hardly
scientifically acceptible.

In conclusion, we have presented 3 fundamentally different electronic
structures for 3d electron states in LaCoO$_3$ discussed in the current
literature. We are in favour of the localized electron atomic-like approach
that yields the discrete energy spectrum associated with the atomic-like
states of the Co$^{3+}$ ions. This approach bridges the atomic physics and
the solid-state physics. It reveals very strong correlations of the local
magnetic moment and the local symmetry (of the crystal field). Our approach
provides in the very natural way the non-magnetic low-temperature state (the
3d$^6$ highly-correlated system is a non-Kramers system) and the insulating
state in LaCoO$_3$. Good description of many electronic and magnetic
properties indicates that the band-structure calculations have to be
oriented into the very strong intra-atomic d-d correlation limit in order to
get the ground state in agreement with two Hund's rules and take into
account the orbital magnetism.

{\bf Acknowledgement}.The authors are grateful to R.Michalski for the
assistance in the computer preparation of this paper.

{\bf Figure caption:}

Fig. 1. Schematic description of the d states in LaCoO$_3$ within the band
approach - there is the continuous energy spectrum.

Fig. 2. Single-electron discrete energy spectrum of the Co$^{3+}$ ion in
LaCoO$_{3}$ for the low-spin (LS, S=0), intermediate-spin (IS, S=1) and
high-spin state (HS, S=2). According to the current literature these spin
states are subsequently realized with the increasing temperature.

Fig. 3. a) The fine electronic structure of the Co$^{3+}$ ion in LaCoO$_{3}$
produced by the octahedral crystal field and the intra-atomic spin-orbit
coupling. b) The lowest part of the fine electronic structure of the Co$^{3+}
$ ion in LaCoO$_{3}$ originating from the cubic subterm $^{5}$T$_{2g}$ with
further splittings by the trigonal distortion. The non-magnetic ground state
as well as the intermediate and high-magnetic states should be noticed.
According to the CEF\ approach these spin states are increasingly populated
with the increasing temperature.

\end{document}